\documentstyle[12pt]{article}
\newcommand{\HU}{\hat{U}}

\newcommand{\HH}{\hat{H}}
\newcommand{\HT}{\hat{T}_\varepsilon}

\newcommand{\HGO}{\hat{G}_\varepsilon^{(0)}}
\newcommand{\half}{\frac{1}{2}}

\newcommand{\shalf}{{\scriptstyle \frac{1}{2}}}
    
\begin{document}%
\title{\Large \bf
Quantum Tunneling in the Wigner Representation}
\author{
{\normalsize\sc M. S. Marinov and Bilha Segev }\\
{\normalsize\it
Department of Physics, Technion-Israel Institute of Technology}\\
{\normalsize\it Haifa 32000, Israel}\\
}

\date{February 1996 }

\maketitle
{\begin{abstract}
Time dependence for barrier penetration is considered in the phase space.
An asymptotic phase-space propagator for nonrelativistic scattering on a 
one - dimensional barrier is constructed. 
The propagator has a form universal for various initial state 
preparations and local potential barriers.
It is manifestly causal and includes time-lag 
effects and quantum spreading.
Specific features of quantum dynamics which disappear in the 
standard semi-classical approximation are revealed. 
The propagator may be applied to calculation of the 
final momentum and coordinate distributions, for particles
transmitted through or reflected from the potential barrier,
as well as for elucidating the tunneling time problem.
\end{abstract}

{\em PACS number}: 03.65.Nk

\section{Introduction}
Observable properties of quantal systems, like energy levels and
transition probabilities are mostly related to stationary states. 
Meanwhile,
the time dependence of physical processes is also described by quantum 
theory and may be of considerable interest. An important class of effects
is various barrier penetration (tunneling) processes. The transition 
probabilities are usually obtained by means of the time-independent 
(energy) methods, in particular, in the semi-classical approximation 
(see e. g. in Ref.\cite{landau-qm,schiff}). As soon as one gets the complete solution 
in the energy representation, the time evolution is obtained 
straightforwardly, in principle, in terms of the inverse Laplace transform.
However, the evaluation of the large-time asymptotics may be an intricate 
job. A source of the trouble is in the very statement of the problem. 
If we insist that the particle was on one side of the barrier in the 
beginning, the state cannot be described by the plane wave, which is the 
eigen-state of the momentum operator. Thus the initial energy is never 
free of an uncertainty. The uncertainty may be made smaller if the 
particle is de-localized in space, so one has to start far enough from
the barrier, and to detect the result long enough after the start in 
order to be sure that the particle has left the potential domain completely.
It is clear, however, that the problem needs a special theoretical 
analysis.

The time dependence of tunneling 
processes has been attracting attention for decades. 
A controversial question is that of the tunneling time and 
the effect of causality on the particle propagation
\cite{buttiker&landauer}-\cite{steinberg}. 
New experimental techniques enable detailed measurements performed 
on a photon wave packet transmitted through an optical analog of 
a quantum potential barrier\cite{ranfagni}-\cite{krausz}.
There are some theoretical concerns on the validity of the 
semi-classical approximation in processes, like tunneling,
which have no classical counterparts, because the effect of quantal
fluctuations must get a proper account. Therefore a consistent 
time-dependent formalism for tunneling is hardly redundant.
 
Barrier penetration processes and their time dependence were investigated 
by numerical, experimental, and analytic methods in a number of 
works, e.g.\cite{winter}-\cite{moshinsky}. The literature contains
specific examples of barriers and wave-packet shapes. Our purpose was
to consider a general case, with no assumptions on the initial state 
preparation, neither on the form of the (local) potential barrier. 
Barrier penetration is described sometimes by means of the
imaginary time method\cite{mclaughlin}-\cite{aoyama}.
In the present approach, 
the process is described in real space-time,
even though we exploit analytical properties in the complex energy plane, 
especially for causality arguments.

The method applied here is an investigation of the time evolution of the 
Wigner phase-space distribution\cite{wigner}. In this way, we can 
consider any initial state, not necessarily pure, which is important for
applications to experiments. Besides,
cumbersome oscillations of the wave functions are not involved. 
The Wigner function was used successfully in many problems of 
quantum theory, and its properties where considered, e.g. in
\cite{carruthers}-\cite{mrowczynski}. 
Carruthers and Zachariasen, in their review of quantum
collision theory with phase-space distributions\cite{carruthers} 
considered  cross sections for scattering
processes in three dimensions, inclusive reactions within a
second quantization approach, inclusive multi-particle
production processes in the ultrarelativistic domain and many
other processes but discussed neither quantum jumps, nor barrier penetration.
Various other approaches have been tried
recently\cite{royer}-\cite{smerzi} to describe quantum dynamics in 
terms of the Wigner function. 
In scattering problems, the time evolution of 
the Wigner function was considered mainly within the 
semi-classical approximation\cite{JACKIW&WOO}-\cite{peev}. 
As was shown by Berry\cite{BERRY}, 
the semi-classical approximation 
for stationary Wigner functions describing bound states is given by 
the Airy function, its spread from the classical $\delta$-function being
the first-order quantum effect. 
Propagation of wave packets was discussed 
in a number of works; it was shown in particular that 
the spreading 
in the coordinate is not a specific quantal effect\cite{LITTLEJOHN}. 
Within the semi-classical approximation 
the quantal features of the
long-time evolution was attributed to the interference between amplitudes 
corresponding to different classical paths\cite{stroud}.
Our point is to emphasize the difference between the genuine
quantum dynamics and the semi-classical approximation for 
classically forbidden processes where no classical paths exist, 
so quantum dynamics is not reduced just to a smearing around 
classical paths, or to an interference between them. 

The Wigner function was applied to tunneling by Balazs and 
Voros\cite{balazs&voros} for parabolic potential barrier. 
The equivalence of Wigner's integro-differential equation 
to Liouville's classical equation is in an apparent conflict to
tunneling for that potential.  The puzzle is solved as soon as
it is realized that the initial Wigner function cannot be chosen 
arbitrarily, if the potential does not vanish asymptotically. 
Physically available states correspond to Wigner distributions
extended in the momentum, so real classical trajectories transport the 
particle above the barrier.  The Wigner function for the parabolic
potential was constructed explicitly, and the transmission and
reflection coefficients were obtained, within a qualitative picture of
tunneling where the difference between classical and quantum mechanics
lies mainly in the initial state preparation. 

For sufficiently broad incident momentum distributions,
the classical paths enable the particle to overcome the barrier, and 
the semi-classical approach can be used indeed\cite{muga-num,defendi}. 
If the barrier potential is localized, the incident wave packet (or the Wigner
function) can be prepared with an arbitrarily definite energy, 
and no classical paths would be responsible for the tunneling.  
In that situation the barrier penetration must be a result of an 
essential difference between the quantum and classical dynamics. 

Purely quantal effects and their role in scattering processes
have been also considered in the phase space formalism. 
In the ``Wigner-trajectory'' approach each phase-space point in the initial 
Wigner distribution is propagating along a definite 
trajectory\cite{muga&brouard&sala,scullyfound,jensen} . 
If the third and higher-orders derivatives of the potential vanish,
the Wigner trajectories coincide with the classical paths. 
Otherwise, the Wigner trajectories are defined with a 
modified ``quantum'' potential and are not classical.
For a system in an energy eigen-state
(i.e. in the stationary barrier problem), 
the time-shift invariance implies that the trajectories 
are the ``equi-Wigner curves'' which are lines of constant values of 
the Wigner function. That approach seems rather problematic. 
The effective `potential' may be singular, the
Liouville theorem is violated, and quantum jumps\cite{remler,garraway} 
can hardly be included.
In another approach, quantum corrections to classical dynamics are
interpreted as finite momentum jumps between classical paths\cite{remler}. 
Negative quasi-probabilities appear in the the calculation, which
distinguish the quantum treatment from the classical theory.
The phase-space points are smeared to finite domains and do not
propagate along continuous trajectories. 

We consider the phase-space evolution kernel\cite{marin91}, which is the
fundamental solution of the dynamical equation for the Wigner function.
In classical theory, the evolution kernel is
the fundamental solution of the Liouville equation and equals 
the $\delta$-function restricted to classical trajectories. 
In the semi-classical approximation one can get the Airy function.
For the barrier penetration, an explicit expression is obtained 
and it is shown that it is not reduced to the semi-classical 
approximation. From this point of view, 
the barrier penetration is an essentially quantal process.

In Section 2 the phase-space propagator (the evolution kernel) is
defined, and some of its properties are given. Section 3 shows the 
relation between the time evolution and the $S$-matrix formalism in
the momentum representation. The large-time asymptotics for the space-time
propagator is derived in Section 4. The result is an integral representation
in terms of scattering amplitudes.
The exact result for the narrow potential barrier is presented in Section 5.
The semi-classical approximation is considered in Section 6, and its validity
is discussed. Some technical aspects are considered in Appendix:
the accuracy of the large-time asymptotics, the transition probability
for Gaussian states, and the exact result for the 
$\cosh^{-2}$ potential barrier.

The units used in the paper are $\hbar=1$, and $2m=1$ for the particle mass.

\section{The phase-space propagator}
In general,
any quantum state is described by a density matrix $\hat{\rho}$,
which can be represented by its matrix elements, say, in the coordinate
representation, $\langle q'|\hat{\rho}|q\rangle$, or by its Weyl symbol 
(the Wigner function\cite{wigner}) $\rho(x)$, where $x\equiv(q,p)$, 
\begin{equation}   
\rho(q,p)=\int^\infty_{-\infty}
\langle q+\frac{\eta}{2}|\hat{\rho}|q-\frac{\eta}{2}\rangle
e^{-ip\eta}d\eta.                                         
\end{equation}
For a given Hamiltonian $\HH$, the
evolution operator $ \hat{U}(t)\equiv\exp(-i\hat{H}t)$
determines the time evolution of the state:
 $\hat{\rho}_t = \hat{U}(t)\hat{\rho}_0\hat{U}^\dagger(t)$.
Respectively, the time evolution of the Wigner function is given by 
an integral kernel, i.e. the phase-space propagator,
     \begin{equation}
 \rho_t(x) = \int L_t(x,x_0)\rho_0(x_0)dx_0.   
      \end{equation}
Here $dx=dqdp$, and $L_t$ is a real dimensionless function 
which satisfies the following identities,
   \begin{eqnarray}
&&L_t(q,p;q_0,p_0)=L_{-t}(q_0,-p_0;q,-p),\nonumber\\
&&\int dx L_t(x,x_0)= 1 =\int dx_0 L_t(x,x_0),   \nonumber \\
&&L_{t_1+t_2}(x,x_0)=\int dx^\prime L_{t_2}(x,x^\prime)
L_{t_1}(x^\prime,x_0) 
    \end{eqnarray}
In order to handle experimental data, one may introduce an 
{\em acceptance function} $\zeta(x)$, characterizing the apparatus,
so that the probability to detect the system described by $\rho(x)$ would be
        \begin{equation}      
w_\zeta=\int\zeta(x)\rho(x)dx.    \end{equation}
Thus the probability to detect the system at a time $t$ after it was 
prepared in the state given by $\rho_0(x)$ is
    \begin{equation}      
w_\zeta(t)=\int\!\int\zeta(x)L_t(x,x_0)\rho_0(x_0)dx_0dx.
   \end{equation}
The quasi-distributions $\rho(x)$ and $\zeta(x)$ are normalizable
and satisfy a set of
conditions owing to the positive definiteness of the density matrix, 
for instance,
    \begin{equation}      
\int[\rho(x)]^2dx\le\left[\int\rho(x)dx\right]^2.
   \end{equation}
Other conditions are more intricate. Qualitatively, the distributions 
cannot be localized to domains of areas less than $2\pi\hbar$ by the 
order of magnitude. The functions are not necessarily positive
everywhere, but the domains of negativity must be small enough.

The {\em phase-space propagator} $L_t$ is expressed in terms of the matrix
elements of $\hat{U}(t)$, e.g. the coordinate (or momentum) propagators, 
as follows from (1), 
     \begin{eqnarray}
L_t(q,p;q_0,p_0) =\frac{1}{2\pi}\int^\infty_{-\infty} d\eta
\int^\infty_{-\infty} d\eta_0 e^{i(p\eta-p_0\eta_0)}
\langle q-\frac{\eta}{2} |\hat{U}|q_0-\frac{\eta_0}{2} \rangle
\overline{\langle q+\frac{\eta}{2}
|\hat{U}|q_0+\frac{\eta_0}{2}\rangle}\nonumber \\
\equiv\frac{1}{2\pi}\int^\infty_{-\infty} d\sigma
\int^\infty_{-\infty} d\sigma_0 e^{i(q\sigma-q_0\sigma_0)}
\langle p + \frac{\sigma}{2}|\hat{U}(t)|p_0+\frac{\sigma_0}{2}\rangle
\overline{\langle p-\frac{\sigma}{2}
|\hat{U}(t)|p_0-\frac{\sigma_0}{2}\rangle}.
\end{eqnarray}

For any Hamiltonian, quadratic in $x$, $\rho_t(x)$ is the solution of 
the {\em classical} Liouville equation. In that case the classical equations 
of motion are linear, as well as the Heisenberg equations. The solution
is linear: $x=R_tx_0$, where $R_t$ is an $x$-independent matrix, and
$L_t(x;x_0) =\delta(x-R_tx_0)$. 
In particular, for the nonrelativistic free motion one has
   \begin{equation}
L_t(q,p;q_0,p_0) =\delta(p-p_0)\delta(q-vt-q_0), 
    \end{equation}
where $v=p/m$ is the particle velocity.

For non-linear systems, effects specific for quantum dynamics result
in deviations of $L_t$ from the $\delta$-function.

\section{Time dependence in the momentum representation}
As soon as the Hamiltonian $\hat{H}$ has no explicit time dependence, 
the evolution operator can be written as the Laplace transform of 
the resolvent:
     \begin{equation}
\hat{U}(t)\equiv e^{-it\hat{H}} =
\frac{1}{2\pi i}\int_{\Gamma_\infty} \hat{G}_\varepsilon
e^{-it \varepsilon    } d\varepsilon \ \ \ , \ \ \ \          
\hat{G}_\varepsilon\equiv(\hat{H}-\varepsilon)^{-1} \ .       
      \end{equation}
where $\Gamma_\infty$ is the usual integration contour in the complex
$\varepsilon$-plane, running above the real axis.

We shall consider a nonrelativistic particle of mass $m=1/2$ scattered 
from a localized one-dimensional and time-independent potential barrier.  
The Hamiltonian is $\hat{H}=\hat{H}_0+\hat{V} $, where $\hat{H}_0=\hat{p}^2$ 
is the kinetic energy operator. The basis of the normalized momentum 
eigen-states $|k\rangle$ will be used, so
     \begin{eqnarray}
\hat{H}_0|k\rangle = k^2 |k\rangle,\;\;\;
\;\;\; \langle k|k_0\rangle=\delta(k-k_0).
        \end{eqnarray}
Introducing the transition operator $\hat{T}_\varepsilon$, one has
\begin{equation}
\hat{G}_\varepsilon =
\hat{G}_\varepsilon^{(0)} -
\hat{G}_\varepsilon^{(0)} 
\hat{T}_\varepsilon 
\hat{G}_\varepsilon^{(0)}   \ \ \ , \ \ \ \    
\hat{G}_\varepsilon^{(0)} \equiv
(\hat{H}_0-\varepsilon)^{-1}.    
\end{equation}
The momentum propagator is now given by
\begin{eqnarray}
&&\langle k|\HU(t)|k_0\rangle =  
\frac{1}{2\pi i} \int_{\Gamma_\infty} \langle k | \hat{G}_\varepsilon
|k_0\rangle e^{-it \varepsilon    }d\varepsilon   \\
&&= \delta(k-k_0)e^{-itk^2}-
\frac{1}{2\pi i} \int_{\Gamma_\infty}d\varepsilon e^{-it \varepsilon    }
\frac{ \langle k|\hat{T}_\varepsilon|k_0\rangle }
{(k^2 - i\gamma-\varepsilon)
(k_0^2 - i\gamma-\varepsilon) } 
\nonumber
\end{eqnarray}
The infinitesimal positive quantity $\gamma$ is introduced here to specify 
the integral near the poles due to the free propagators.
The integration contour $\Gamma_\infty$ is deformed to  
run around the positive real axis, 
as the exponential vanishes in the lower half-plane and
$\langle k|\hat{T}_\varepsilon|k_0\rangle$ has singularities 
only for the real values of $\varepsilon$ corresponding to physical 
energy values. The kinematic poles are isolated, leaving an integral along
the real positive axis 
     \begin{eqnarray}
\langle k|\HU(t)|k_0\rangle = \exp[-\frac{i}{2}t(k^2+k_0^2)]
\left[\delta(k-k_0)-\frac{e^{it\xi} }{2\xi}
\langle k|\hat{T}_\varepsilon|k_0\rangle \mid_{\varepsilon=k^2 }
\right.\\ 
+\left.\frac{e^{-it\xi} }{2\xi}
\langle k|\hat{T}_\varepsilon|k_0\rangle \mid_{\varepsilon=k_0^2 } 
- J_t(k.k_0)\right],        \nonumber
\end{eqnarray}
where $\xi=\frac{1}{2}(k_0^2-k^2)$ and
         \begin{equation}   
J_t(k,k_0) = \frac{1}{\pi}{\rm p.v.} \int^\infty_{\varepsilon_0} 
d\varepsilon e^{-it[\varepsilon-\frac{1}{2}(k^2+k_0^2)]    }
\frac{{\rm Im} [\langle k|\hat{T}_\varepsilon|k_0\rangle] }
{(\varepsilon-k^2)(\varepsilon-k_0^2)},  
  \end{equation}
and $\varepsilon_0$ is a threshold energy value. (With no bound states
$\varepsilon_0=0$, otherwise it is the lowest bound-state energy.)
At two zeroes of the integrand denominator
the integral is taken in the sense of its principal value. 

In order to calculate explicitly the time-dependent propagator for
a given potential, the transition operator matrix elements for this
potential should be known on and off the energy shell. 
For scattering problems, one needs the large-time asymptotics,
where the matrix elements are reduced to the energy shell
because of the known fact of the theory of distributions,
     \begin{equation}   
\lim_{t\rightarrow\infty} e^{i\xi t}/\xi = i\pi\delta(\xi).
      \end{equation}
The integral in (14) vanishes, as $t\rightarrow\infty$, because
 Im $[\langle k|\hat{T}_\varepsilon|k_0\rangle]$ is smooth 
and the integral is converging. In general,
$J_t=$ O$(t^{-1/2})$.
(The proof is given in Appendix A.)

The result is expressed in terms of a unitary $2\times 2$ matrix $S$
     \begin{equation}  
\langle k|\hat{U}(t)|k_0 \rangle\asymp e^{-i\kappa^2 t}2\kappa
\delta(k^2-k_0^2)S_{\nu\nu_0},\;\;\;
SS^\dagger=I,
        \end{equation}
where $\kappa=|k|=|k_0|$, $I$ is the unit $2\times 2$ matrix and 
$\nu=k/\kappa=\pm 1$. 
The $S$-matrix elements are related to elements of the $\HT$-operator on
the energy shell $\kappa^2\equiv\varepsilon$,
   \begin{equation}
S_{\nu\nu_0}(\kappa)=\delta_{\nu\nu_0}-\frac{i\pi}{\kappa}
\langle\nu\kappa|\HT |\nu_0\kappa\rangle.
    \end{equation}
These are the probability amplitudes for transmission through and
reflection from the potential region. The amplitudes can be expressed in 
terms of two analytical functions $a(\kappa)$ and $b(\kappa)$,
  \begin{equation}    
S(\kappa)=\frac{1}{a}\left(\begin{array}{cc}
1 & b\\
-\bar{b} & 1
\end{array}\right),\;\;\;|a|^2-|b|^2=1.
  \end{equation} 
These functions are defined for Re $\kappa>0$ by the asymptotics of the
solution of the stationary Schr\"{o}dinger equation,
\begin{eqnarray}   
\hat{p}^2y+V(q)y=\kappa^2y,    \nonumber\\
y_-(q)\asymp\left\{\begin{array}{cc}
e^{-i\kappa q}, & q\rightarrow-\infty,\\
ae^{-i\kappa q}+be^{i\kappa q}, & q\rightarrow +\infty.
\end{array}\right.  \end{eqnarray}
The functions $a(\kappa)$ and $b(\kappa)$ have the analytical continuation 
to the left half-plane by
  \begin{equation}
a(-\bar{\kappa})=\overline{a(\kappa)},\;\;\;
b(-\bar{\kappa})=\overline{b(\kappa)}.
   \end{equation} 
The analytical properties of these functions in the complex $\kappa$-plane
have been investigated previously\cite{analy}. 
It was shown, in particular, that for any finite-range 
and positive potential 
they can be expressed in terms of two entire
functions $\alpha(\varepsilon)$ and $\beta(\varepsilon)$,
  \begin{equation}
a(\kappa)\equiv 1-\alpha(\varepsilon)/2i\kappa,\;\;\;\;
b(\kappa)\equiv\beta(\varepsilon)/2i\kappa.
   \end{equation} 
Thus the only singularities of $S(\kappa)$ are poles
due to zeroes of $a(\kappa)$ which are all in the lower half of the
$\kappa$-plane. Moreover, $\alpha$ and $\beta$
are bounded for Im $\kappa>0$, so that lim $a=1$ and lim $b=0$, as 
$|\kappa| \rightarrow\infty$ in the upper half-plane. If the potential 
has an exponential decrease as $q\rightarrow\pm\infty$, the functions 
$\alpha$ and $\beta$ may have infinite series of poles. 
Besides, for symmetric barriers, where $V(-q)=V(q)$, one has a real 
$\beta(\varepsilon)$ and purely imaginary $b(\kappa)=-b(-\kappa)$.

\section{The large-time asymptotics}
In order to get the large time asymptotics for the phase-space propagator, 
we set the amplitudes from (16) into Eq. (6). Using the following property
of the $\delta$-function
     \begin{equation}  
2|k|\delta(k^2-k_0^2)=\delta(k-k_0)+\delta(k+k_0),
     \end{equation}
one gets the following result
     \begin{eqnarray}  
L_t(q,p;q_0,p_0)\asymp
\delta(p-p_0){\cal T}(p_0,r_+)+\delta(p+p_0){\cal R}(p_0,r_-) \\
+ \frac{2}{\pi}{\rm Re}\left[
\frac{b(p_0-p)}{a(p_0-p)a(p_0+p)} 
e^{2i(q_0p-qp_0)+4ipp_0t}\right] . \nonumber
     \end{eqnarray}
Here $r_\pm=q_0+2p_0t\mp q$; i.e the differences between the free 
classical trajectory and actual positions of the transmitted particle 
($r_+$) and reflected from $q=0$ particle ($r_-$). The first two terms 
describe transmission and reflection from the potential barrier. 
The third term represents an interference between transmitted and reflected
waves and is responsible for quantum fluctuations at the barrier region.
It is irrelevant for large $t$, if the wave packet was prepared
in free space with a narrow momentum distribution.

The functions representing the transmission and reflection probabilities are 
given by the Fourier integrals
      \begin{eqnarray}
 {\cal T}(p_0,r_+)=\frac{1}{2\pi}\int^{+\infty}_{-\infty}
\frac{e^{-i\sigma r_+}d\sigma}{a(\half\sigma+p_0)a(\half\sigma-p_0)}, \\
 {\cal R}(p_0,r_-)=\frac{1}{2\pi}\int^{+\infty}_{-\infty}
\frac{b(\half\sigma+p_0)b(\half\sigma-p_0)e^{-i\sigma r_-}d\sigma}
{a(\half\sigma+p_0)a(\half\sigma-p_0)}\\
+\frac{1}{2\pi}\int^{+\infty}_{2p_0}\frac{B(p_0,\sigma)d\sigma}
{a(\half\sigma+p_0)a(\half\sigma-p_0)},\nonumber
    \end{eqnarray}
where 
\begin{equation}
B(p_0,\sigma)\equiv[b(\shalf\sigma+p_0)+b(-\shalf\sigma-p_0)]
[b(\shalf\sigma-p)e^{-i\sigma r_-}+b(-\shalf\sigma+p)e^{i\sigma r_-}].
\end{equation}
The result is obtained in the following way. We assume that 
$p_0>0$, so $L_t$ gets
 contributions from 3 segments in the $\sigma$-axis, which
are proportional to products of the $S$-matrix elements, as follows:
\begin{equation}    
\begin{array}{lccc}
\sigma\in &(-\infty,-2p_0) & (-2 p_0,2 p_0) & (2 p_0,+\infty)\\
  &  &  &  \\ 
{\cal T}\propto & S_{--}\overline{S_{++}} & S_{++}\overline{S_{++}} & 
S_{++}\overline{S_{--}}\\
  &  &  &  \\ 
{\cal R}\propto & S_{-+}\overline{S_{+-}} & S_{+-}\overline{S_{+-}} &
S_{+-}\overline{S_{-+}}
\end{array}
\end{equation}
The arguments are $p+\half\sigma$ in the first factor and $p-\half\sigma$
in the second factor. The $S$-matrix is given in Eq. (18), and the complex
conjugation is taken on the real axis by means of Eq. (20), inverting 
the signs of the arguments. Thus the integral is given in terms of 
analytical functions with the general properties determined by the 
Schroedinger equation.

Evidently, $L_t$ is real and respects the reciprocity principle:
   \begin{equation}   
L_t(x,x_0)=L_{-t}(x_0,x).
\end{equation}
Note that the result is translationally invariant; namely, for 
$q\rightarrow q-c$ the quantities $r_+$ and $a(\kappa)$ remain invariant, 
while $r_-\rightarrow r_--2c$ and
$b(\kappa)\rightarrow b(\kappa)e^{2i\kappa c}$.
The second integral in $\cal R$ vanishes, if $V(q)\equiv V(-q)$,
since in that case $b(\kappa)=-b(-\kappa)$ and $B(p,\sigma)\equiv 0$. 
The total transmission and reflection probabilities 
(for a given initial momentum $p$) are given by integration
       \begin{eqnarray}    
T(p)=\int^{+\infty}_{-\infty}{\cal T}(p,r)dr=\left|a(p)\right|^{-2}, \\
R(p)=\int^{+\infty}_{-\infty}{\cal R}(p,r)dr=\left|b(p)/a(p)\right|^2.\nonumber
       \end{eqnarray}

The integral representations in Eqs. (24-25) are manifestly causal.
It is proven\cite{analy} that the only singularities are due to 
zeroes of $a(\kappa)$, which are all in the lower complex 
half-plane for $V(q)\ge 0$. 
If $r_+< 0$, i.e. if $q$ is ahead of the free propagation coordinate, 
the exponent for Im $\sigma>0$ is decreasing, so the integration contour
may be deformed to the upper half plane and the only contribution would be
from $\infty$ where $a\rightarrow 1$, leading to ${\cal T}=\delta(r_+)$,
as in free motion (8). For $r_+\ge 0$, the integral can be evaluated by
means of residues in the lower half plane. Each zero of $a(\kappa)$ gives
rise to a pair of poles in the integrand of (24) and (25). Thus 
${\cal T}$ and ${\cal R}$ can be expressed as a sum over 
the $S$-matrix singularities. The transmission propagator
is given by,
     \begin{equation}   
{\cal T}(p,r_+)=\delta(r_+)-\theta(r_+)\Sigma_{n}{\rm Re} 
\left\{A_{n}(p)\exp[ i 2 r_+ (p-\kappa_n) ]\right\},
      \end{equation}
where $\kappa_n$ are zeroes of $a(\kappa)$ and $A_n(p)$ are 
determined by the corresponding residues,
$\theta(r)$ is the step function ($=0$ for $r<0$ and $=1$ for
$r>0$). Two examples are given below: the $\delta$-potential barrier 
(Section 5) and the modified P\"{o}schl-Teller potential 
(Appendix C). In general, contributions from purely imaginary poles
have the form of that for the $\delta$-barrier, Eq. (33).
The functional dependence on the lag distance $r_+$ is universal, 
though $A_n$ and $\kappa_n$ depend on the potential. As  
Im $\kappa_n < 0$, the second term in ${\cal T}(p,r_+)$ is an exponentially 
decreasing and oscillating function of $r_+$. It partly cancels 
contributions from the $\delta$-function to the integrals for probabilities, 
so the effect of the barrier is to remove
parts of the freely propagating wave packet.

\section{Narrow potential barrier} 
Let us consider the $\delta$-potential barrier
      \begin{equation}    
V^{\delta}(q)=v_0\delta(q),
\end{equation}
for which the $S$-matrix elements are given by
    \begin{equation}    
a^{\delta}(\kappa)=1-v_0/2i\kappa,\;\;\;b^{\delta}(\kappa)=v_0/2i\kappa.
    \end{equation}
There is one pair of poles in the integrands in (24) and (25), and the 
integrals are calculated by the residue theorem, leading to
     \begin{eqnarray}   
{\cal T}^{\delta}(p,r_+)=\delta(r_+)-\theta(r_+)
2v_0\sqrt{1+(v_0/4p)^2}e^{-v_0r_+}\cos(2pr_++\gamma)\\
{\cal R}^{\delta}(p,r_-)=
\theta(r_-)\frac{v_0^2}{2p}e^{-v_0r_-}\sin(2pr_-),\nonumber 
      \end{eqnarray}
where $\gamma=\arctan (v_0/4p)$. 
The total transmission probability is
$$ T(p)=1-\frac{v_0^2}{v_0^2+4p^2}.$$

The $\delta$-potential has a simple explicit solution also off the energy 
shell, 
       \begin{equation}   
\langle k|\HT^{\delta} |k_0\rangle =
\frac{v_0}{2\pi(1+v_0/2\sqrt{-\varepsilon})},
       \end{equation}
independently of $k$ and $k_0$. Thus one can evaluate the non-asymptotic
time dependence of Eq. (14),
       \begin{eqnarray}  
J^{\delta}_t(k,k_0)=\frac{v_0^2}{4\pi\sqrt{-it}}e^{\frac{i}{2}t(k^2+k_0^2)}
\left[\frac{W(\shalf v_0\sqrt{-it})}
{(k^2+\frac{1}{4}v_0^2)(k_0^2+\frac{1}{4}v_0^2)}\right.\\
\left.+\frac{1}{k^2-k_0^2}\left(
\frac{W(\sqrt{itk^2})}{k^2+\frac{1}{4}v_0^2}-
\frac{W(\sqrt{itk_0^2})}{k_0^2+\frac{1}{4}v_0^2}\right)\right],\nonumber
       \end{eqnarray}
where $W(z)$ is a function related to the probability integral\cite{abram} 
       \begin{equation}  
W(z)=zw(z)\equiv ze^{-z^2}\left(1-{\rm erf}(-iz)\right)\asymp
\frac{i}{\sqrt{\pi}}\left(1+\frac{1}{2z^2}+{\rm O}(z^{-4})\right)
       \end{equation}
 In the large-time asymptotics, 
for $t\gg\max\left\{ v_0^{-2}, k_0^{-2}, k^{-2}\right\}$,
$J^{\delta}_t$ vanishes as $t^{-3/2}$, more rapidly than in the general case
since the transition matrix elements in Eq. (34) are insensitive to 
separation from the energy shell (cf. Appendix A). The result is
       \begin{equation}  
J^{\delta}_t=\frac{\exp[\frac{i}{2}(k^2+k^2_0)t]}
{2(i\pi t)^{3/2}k^2k^2_0}+O(t^{-5/2}).
       \end{equation}
As can be shown by means of the double Fourier transform in $k$ and $k_0$,
the expression in Eq. (35) agrees with the evolution kernel for the 
$\delta$-potential in the coordinate representation, 
obtained previously\cite{schulmandelta}. 

\section{Semi-classical approximation}
In the semi-classical approximation one has the eikonal formula,
for the transmission amplitude
    \begin{eqnarray}    
a(\kappa)=\exp [iS(\kappa)],\;\;\;
S(\kappa)=\kappa\int^{+\infty}_{-\infty}
\left[1-\sqrt{1-\kappa^{-2}V(q)}\right]dq.
    \end{eqnarray}
For $\kappa^2>V_0\equiv\max[V(q)]$, $S$ is real and $|a|=1$, so the 
transmission is complete. Note that $S(-\kappa)=-S(\kappa)$, by virtue of 
the analytical continuation. Unlike the exact solution, the approximate 
function $a(\kappa)$ is not meromorphic, since $S(\kappa)/\kappa$ has a 
branch point at $\kappa^2=V_0$. 

The resulting transmission propagator is given by
    \begin{eqnarray} 
{\cal T}(p,r)= \frac{1}{\pi} \int^{\infty}_{0}d\sigma
\cos\left[\sigma r +S(p+\frac{\sigma}{2})-S(p-\frac{\sigma}{2})\right].
    \end{eqnarray}

The total transmission probability is obtained by integrating 
$\cal T$ in $r_+$. At large $r_+$ the vicinity of $\sigma=0$ 
is dominating in the integral, so one can use the series expansion 
in the exponent. For  $p^2<V_0$,
    \begin{equation}    
S(p)+S(-p)=-2iI,\;\;\;I\equiv\int\sqrt{V(q)-p^2}dq,
    \end{equation}    
where the integral is on the segment where $V(q)>p^2$, $I$ is real. 
That leads to the familiar semi-classical expression for the transmission 
probability, $T=\exp(-2I)$.

Unlike the integrated probability, the value of the propagator for a 
given $r_+$ needs a more careful treatment.
Expanding the exponent around $\sigma=0$ and retaining terms up to the 
order of $\sigma^3$, one gets the Airy function for ${\cal T}(p,r_+)$. 
While adequate for classically allowed regions\cite{BERRY}, 
as well as for large $r_+$, this naive semi-classical approximation can 
not be valid everywhere. In particular, it violates casuality in 
contrast to the results obtained from the exact quantum treatment.
 
Fairly accurate results in some domains in the $(p_0,r_+)$-plane
are obtained in the framework of the WKB approach if the integral in Eq. 
(39) is calculated by the stationary-phase method. 
If the critical points of the integrand appear in the region where  
the semi-classical approximation is valid, one can make use of Eq. (38).
Positions of the critical points are given by the equation 
    \begin{eqnarray}    
2r_+ +S'(\shalf\sigma+p)+S'(\shalf\sigma-p)=0,\\
S'(\kappa)=\int^{+\infty}_{-\infty}
\left[1-\frac{\kappa}{\sqrt{\kappa^2-V(q)}}\right]dq.\nonumber
    \end{eqnarray}
The latter has a classical meaning for $\kappa^2>V_0$; namely, 
$S'(\kappa)=-v(\tau_V-\tau_0)$, where $v\equiv 2\kappa$ is the initial 
particle velocity, $\tau_V$ and $\tau_0$ are the times of flight in 
presence of the potential barrier and in free space. Hence Eq. (39) can be
interpreted in the following way. Treating $\sigma$ as a quantum 
fluctuation of the incident momentum, one can note that $r_+$ 
equals to the lag due to the barrier, averaged between two classical
trajectories, their mean momentum being equal to $p$.
For positive $r_+$, the critical points of the exponent may take place
at real $\sigma$, and their positions depend on the magnitudes of $r_+$ 
and $p$.

At small $r_+$ the region of large $|\sigma|$ contributes substantially,
and one use the high-energy asymptotics, where
$\kappa\rightarrow\infty$, $a(\kappa)\asymp 1$, and
    \begin{equation}    
S(\kappa)\asymp\frac{w}{2\kappa},\;\;\;\;
w\equiv\int^{+\infty}_{-\infty}V(q)dq.
     \end{equation}
Let us consider the domain where
    \begin{equation}    
r_+p\geq 1,\;\;\;p/w\ll 1,\;\;\;r_+p^2/w\ll 1.
     \end{equation}
The inequalities mean that i) the domain is consistent with the uncertainty
relation, ii) the process is a deep under-barrier tunneling, iii) the 
asymptotics of Eq. (40) can be used. The corresponding critical points are
given by
    \begin{equation}    
\sigma^2_0=2w/r_+\gg p^2.
     \end{equation}
The contribution from the pair of critical points is
    \begin{equation}    
{\cal T}(p,r_+)=\frac{1}{\pi}\int^\infty_0
\cos\left(r_+\sigma+\frac{2w}{\sigma}\right)d\sigma\asymp
\frac{(2wr_+)^{1/4}}{r_+\sqrt{\pi}}
\cos\left(2\sqrt{2wr_+}+\frac{\pi}{4}\right).
     \end{equation}
This result is quite different from the standard semi-classical 
approximation, which is hardly surprising, since the domain given by 
inequalities (43) is not dominating in tunneling as a whole, even 
though small $r_+$ represent the most rapid signal transport.


\section{Summary and conclusion}
The phase-space propagator for a general local one-dimensional potential 
barrier has a universal form; it is given by a sum over singularities of 
the $S$-matrix. It conserves energy, is manifestly causal
and shows certain features specific for quantum theory. 

At large times, the initial Wigner function is splited into three
parts: the reflected bundle, the transmitted bundle and a transient 
group which can be neglected if the incident particle was prepared with 
a narrow momentum distribution. The resulting probabilities have a 
momentum dependence, which causes the known effects, like dispersion and 
forward attenuation. Asymptotically, after the particle leaves the 
interaction region, the time evolution of the Wigner functions is just 
the coordinate translation with constant classical velocities.
As different parts of the Wigner function are translated with  different 
velocities, the evolution of the phase-space distribution
goes on, and the actual experiment results may depend on the detector 
position, in particular on its distance from the barrier region.

The coordinate dependence of the propagators is universal 
for any local potential. The propagators are functions of the lag
distance, which is the difference between the free motion and the 
current coordinate. At large lag distances, the probability has an
exponential decrease and is oscillating as a result of quantum interference
effects. The decrease rate and the oscillation frequency 
are determined by positions of poles of the transition amplitudes in the
complex energy plane. The coefficients and the phase shifts depend on details 
of the barrier shape and on the initial momentum. Of course, the phase-space
propagator is not a $\delta$-function. Each point in the initial Wigner 
function gives rise to disconnected domains in the final distribution. 
This is a purely quantal effect in the barrier penetration dynamics. 

The standard semi-classical approximation can be used for evaluation of the
total transmission probability, obtained by integration of the wave packet
outside the potential domain. It is hardly adequate, however, for description
of the process time dependence, since the analytical properties representing
the causality in the energy representation are not maintained properly.

In contrast to classical theory, individual phase-space trajectories
cannot be traced in quantum theory. Both the the initial distribution and 
the detector acceptance have final supports, and the observed probabilities
are obtained by integration. It is evident, however, that the principle of
causality is not violated by quantum theory, and no information can be 
transported faster because of the barrier. The problem of the signal 
transport and the time delay is a subject of a forthcoming paper\cite{time}.

{\em Acknowledgements}.
We are grateful to S. A. Gurvitz for his interest
to this work. The support to the research from G. I. F.
and the Technion V. P. R. Fund is gratefully acknowledged.

\section*{Appendix A}
\subsection*{Off-energy-shell corrections in the large-$t$ asymptotics}
The transition operator of Eq. (11) has the following general property
   \begin{equation}    
\HT-\HT^\dagger=\HT[\HGO-(\HGO)^\dagger]\HT^\dagger,
    \end{equation}   
which leads, in particular to the $S$-matrix unitarity. 
Hence one has for the discontinuity of the matrix element in Eq. (14)
   \begin{eqnarray}    
{\rm Im}\langle k|\HT | k_0\rangle=\frac{\pi}{2\kappa}D(\kappa;k,k_0),\\
D(\kappa;k,k_0)=\sum_{\nu=\pm1}
\langle k|\HT | \nu\kappa\rangle\overline{\langle k_0|\HT | \nu\kappa\rangle},
 \nonumber
        \end{eqnarray}   
where $\kappa=\varepsilon^{1/2}$. As a function of $\kappa$, $D$ is regular. 
Besides, it has the following general property of positive definiteness,
   \begin{equation}    
\int^\infty_{-\infty}\int^\infty_{-\infty}D(\kappa;k,k_0)f(k)
\overline{f(k_0)}dk_0dk\ge 0,
    \end{equation}   
for any integrable complex function $f(k)$. 
In particular, $D(\kappa;k,k)\ge 0$ for all $k$.

The ``half-on-shell" transition amplitudes present here are expressed in
terms of solutions of the Schr\"{o}dinger equation (19) $y_\pm(x)$, 
satisfying the complimentary asymptotical conditions, 
as given in Ref.\cite{analy},
          \begin{eqnarray}  
&&\langle k|\HT | \nu\kappa\rangle =
\frac{1}{8\pi\kappa a}\int^\infty_{-\infty}dxe^{-i(k-\nu\kappa)x}
\left\{ (\kappa^2-k^2)(\eta'_+\eta_--\eta_+\eta'_-)\right.\\
&&\left.+V(x)\left[((2+\nu)\kappa-k)y_+(x)e^{-i\kappa x}+
((2-\nu)\kappa+k)y_-(x)e^{+i\kappa x}\right]\right\},\nonumber
         \end{eqnarray}   
where $\eta_\pm(x)\equiv y_\pm(x)-e^{\pm i\kappa x}\rightarrow 0$ for 
$x\rightarrow\pm\infty$. It is evident from the integral representation 
that the transition amplitudes are regular functions of
$\kappa$, and they are proportional to $\kappa$ at $k^2=\kappa^2$.
(Note that $\kappa a$ is finite at $\kappa=0$.)

Finally, the off-energy-shell correction is given by the following equality
(we assume, for simplicity that there are no bound states)
   \begin{equation}  
J_t(k,k_0)=\exp[\frac{i}{2}t(k^2+k^2_0)]\;
{\rm p.v.}\int^\infty_0e^{-it\kappa^2}
\frac{D(\kappa;k,k_0)d\kappa}{(\kappa^2-k^2)(\kappa^2-k_0^2)}
\asymp O(t^{-1/2}).
    \end{equation}   
The large-$t$ asymptotics of the integral is determined by the behavior 
of $D$ at $\kappa\rightarrow 0$. 

\section*{Appendix B}
\subsection*{ Gaussian distributions }
Let us consider, for example, the Gaussian initial phase-space distribution
$\rho_0(q,p)$ and a similar detector acceptance function $\zeta(q,p)$.
We shall assume, for simplicity, that they describe pure states which
have the minimal uncertainties, allowed by the Heisenberg principle.
The functions are
     \begin{eqnarray}  
\rho(q,p)= \exp [-(q-Q_0)^2/\lambda_0 -\lambda_0(p-P_0)^2]~~~, \\   
\zeta(q,p)= \exp [-(q-Q_d)^2/\lambda_d -\lambda_d(p-P_d)^2]~~~,\nonumber
     \end{eqnarray}
where $(Q_0,P_0)$ and $(Q_d,P_d)$ are the most probable 
coordinate and momentum values for the initial wave packet and the detector, 
respectively; $\lambda_0$ and $\lambda_d$ are the coordinate dispersions. 
If the wave packet was prepared with a fairly definite initial momentum 
$P_0$, then $\lambda_0$ must be relatively large, 
while $\lambda_d$ is small for a detector localized in space.

The probability to detect the particle at a time $t$ is given by Eq. (5).
For pure states the probability is expressed in terms of the transition 
amplitude. 
    \begin{equation}      
w_\zeta(t)= \frac{1}{2}(\lambda_0\lambda_d)^{1/2}
\exp\left[-\frac{\lambda_0\lambda_d}{\lambda_0+\lambda_d}
(P_0^2+P_d^2)\right]|A+B|^2,
   \end{equation}
where
     \begin{eqnarray}  
A=\int^\infty_{-\infty}\frac{d\kappa}{a(\kappa)}
\exp\left[-\half(\lambda_0+\lambda_d)(\kappa-P_+)^2
-i\kappa(Q_-+\kappa t)\right],\\
B=\int^\infty_{-\infty}\frac{d\kappa}{a(\kappa)}b(\kappa)
\exp\left[-\half(\lambda_0+\lambda_d)(\kappa-P_-)^2
-i\kappa(Q_++\kappa t)\right],\nonumber\\
P_\pm=(\lambda_0P_0\pm\lambda_dP_d)/(\lambda_0+\lambda_d),\;\;\;
Q_\pm=Q_0\pm Q_d. \nonumber
     \end{eqnarray}
Apparently, if the $\kappa$-dependence of $|a|$ is not strong,
the region of $\kappa\approx P_+$ is the most substantial
in the integral for the transmission amplitude $A(t)$. On the other hand,
the amplitude  vanishes in the large-$t$ asymptotics except
for times $t$ given by the stationary
phase condition, 
     \begin{equation}   
vt=Q_d-Q_0-d\varphi/d\kappa,
     \end{equation}
where $v=2\kappa$ is the velocity in the substantial region, and
$\varphi(\kappa)=\arg(a)$. Similar arguments can be applied to $B(t)$.
(A more detailed discussion is given in Ref.\cite{time}.)

The probability in (52) is a sum of three terms, 
$w_\zeta=w_t+w_r+2w_s$, corresponding to transmission, 
reflection and the transient interference, as in Eq. (23) for $L_t$.
At large times $t$ and the states localized outside the potential region 
(large $Q_0$ and $Q_d$), one of the amplitudes is exponentially smaller
than the other one. As soon as $w_s^2=w_tw_r$, the interference term can 
be neglected. In general one deals with mixed states, where
$w_s^2<w_tw_r$, so the interference is always negligible. 

\section*{Appendix C}
\subsection*{Modified P\"{o}schl-Teller potential barrier }
The modified P\"{o}schl-Teller potential barrier,
       \begin{eqnarray}   
V^{\rm PT}(q)=\frac{v_0^2}{\cosh^2(q/s)},
    \end{eqnarray}
where $v_0$ and $s$ are constants, is an example for a local 
potential barrier which enables an explicit solution and a test
of the various approximations. The known solution\cite{landau-qm} is
      \begin{eqnarray}  
&&a^{\rm PT}(\kappa)=i\frac{[\Gamma(1-is\kappa)]^2}
{s\kappa\Gamma(\half+\omega+is\kappa)\Gamma(\half-\omega-is\kappa)}
\nonumber \\
&&b^{\rm PT}(\kappa)=-i\frac{\cos\pi\omega}{\sinh\pi s\kappa},
          \end{eqnarray} 
where $\omega=\sqrt{\frac{1}{4}-v_0^2 s^2}$; it is real if the barrier is
narrow.

The singularities of the transmission amplitude are simple poles
at $s\kappa=-i(n+\half\pm\omega)$, $n=0,1,\cdots$ .
The phase-space propagator is 
given by an infinite sum of the residues. For large $t$ the result is 
       \begin{eqnarray}   
&&L_t^{\rm PT}(q,p;q_0,p_0) \asymp \delta(p-p_0) {\cal T}^{\rm PT}(r_+,p_0)
+ \delta(p+p_0) {\cal R}^{\rm PT}(r_-,p_0), \\
&&{\cal T}^{\rm PT}(r_+,p_0) =\delta (r_+)+ 
\theta (r_+)\left[ {\cal F}_t(\nu, \omega)+{\cal F}_t(-\nu, \omega) + 
{\cal F}_t(\nu, -\omega) + {\cal F}_t(-\nu, -\omega) \right],  
\nonumber \\
&&{\cal R}^{\rm PT}(r_-,p_0) =\theta (r_-)
\left[ {\cal F}_r(\nu, \omega) + {\cal F}_r(-\nu, \omega) + 
{\cal F}_r(\nu, -\omega) + {\cal F}_r(-\nu, -\omega) \right]   \nonumber \\
&&+\theta (-r_-)\left[{\cal F}_s(\nu,\omega)+{\cal F}_s(-\nu,-\omega)\right]  
\nonumber 
         \end{eqnarray}   
where $\nu\equiv 2p_0s$ and, 
         \begin{eqnarray}
&&{\cal F}_t(\nu,\omega) =
\frac{ 2 \Gamma(2\omega)\Gamma(i\nu)\Gamma(i\nu +2\omega)}
{s\Gamma(-\frac{1}{2} + \omega)\Gamma(\frac{1}{2} +\omega) 
\Gamma(-\frac{1}{2} +i\nu +\omega)\Gamma(\frac{1}{2}+i\nu+\omega)}
\times\nonumber \\
&&\exp[(i\nu+2\omega-1)\frac{r_+}{s}]
{~}_4F_3 \left[ \begin{array}{c}
\frac{3}{2}-\omega,\frac{1}{2}-\omega,\frac{3}{2}-i\nu-\omega,
\frac{1}{2}-i\nu -\omega \\
         1- 2\omega , 1- i \nu , 1- i \nu -2\omega  \end{array} 
         ; \exp(-2\frac{r_+}{s})\right],
\nonumber \\ \nonumber \\
&&{\cal F}_r(\nu,\omega)= 
\frac{-2\Gamma(2\omega)\Gamma(i\nu)\Gamma(i\nu+2\omega) 
\Gamma(\frac{1}{2}-i\nu-\omega)}
{s[\Gamma(\frac{1}{2}+\omega)]^2\Gamma(\frac{1}{2}-\omega) 
\Gamma(-\frac{1}{2}+ \omega)\Gamma(-\frac{1}{2}+i\nu+\omega)}
\times\nonumber \\ 
&&\exp[(i\nu+2\omega-1)\frac{r_-}{s}]{~}_4F_3 \left[ \begin{array}{c}
\frac{3}{2}-\omega,\frac{1}{2}-\omega,\frac{3}{2}-i\nu-\omega,
\frac{1}{2}-i\nu -\omega \\
         1- 2\omega , 1- i \nu , 1- i \nu -2\omega  \end{array} 
         ; \exp(-2\frac{r_-}{s})\right]
\nonumber \\ \nonumber \\ 
&&{\cal F}_s(\nu,\omega)=2v_0^2s \frac{\Gamma(\frac{3}{2}+i\nu+\omega)
\Gamma(\frac{3}{2}+i\nu-\omega)\Gamma(-1-i\nu)} 
{\Gamma(\frac{1}{2}+\omega)\Gamma(\frac{1}{2}-\omega)\Gamma(1+i\nu)}
\times\nonumber\\
&&\exp[(i\nu+2)\frac{r_-}{s}]{~}_4F_3 \left[ \begin{array}{c}
\frac{3}{2}+\omega,\frac{3}{2}-\omega,\frac{3}{2}+i\nu+\omega,
\frac{1}{2} +i \nu -\omega \\
         2,1+i\nu,2+i\nu  \end{array} 
         ; \exp(\frac{2 r_-}{s}) \right] .
   \end{eqnarray}   
The ${}_4F_3$ are the generalized hypergeometric 
functions\cite{slater}, 
    \begin{eqnarray}  
{}_4F_3 \left[ \begin{array}{c}
         \xi_1 ,\xi_2 ,  \xi_3 ,\xi_4 \\
         \lambda_1 , \lambda_2 ,\lambda_3 \end{array} ; \zeta \right]
=\sum_{n=0}^{\infty} \frac{\zeta^n}{n!} 
\frac{(\xi_1)_n (\xi_2)_n (   \xi_3)_n ( \xi_4)_n }
{(\lambda_1)_n (  \lambda_2)_n ( \lambda_3)_n },
\end{eqnarray}   
and $(\xi)_n\equiv\Gamma(\xi+n)/\Gamma(\xi)$.
The series are convergent, as long as $|\zeta|<1$. 

For a wide barrier, $\omega$ is imaginary, and the series
approach their radius of convergence as $s\rightarrow\infty$.
Thus the integral representation is more practical than the series.
For small momenta, $\kappa\ll v_0$, one gets:
    \begin{eqnarray} 
a^{\rm PT}(\kappa)\approx\exp[-\pi s(v_0-\kappa)+iS(\kappa)],\\
S(\kappa)=s\kappa\ln(\frac{v_0^2}{\kappa^2}-1)
+sv_0\ln\frac{v_0+\kappa}{v_0-\kappa},\nonumber
\end{eqnarray}   
which is the semi-classical approximation (the integral in Eq. (38) is 
calculated). Now ${\cal T}$ is given by the Airy function,
\begin{eqnarray}
{\cal T}^{\rm PT}_{sc}\approx e^{-2\pi s(v_0-p_0)}
(3\varphi_3)^{-1/3} Ai \left[ (3 \varphi_3 )^{-1/3}(r_+ -\varphi_1)\right],\\
\varphi_1=-s\ln\left(\frac{v_0^2}{p_0^2}-1\right),\;\;\; 
\varphi_3 = \frac{sv_0^2(v_0^2-3p_0^2)}{12p_0^2(v_0^2-p_0^2)^2}.
\nonumber\end{eqnarray}   
The appearance of the Airy function is typical for semi-classical Wigner's 
functions. As $s\rightarrow\infty$, the Airy function approaches the 
$\delta$-function, corresponding to the classical propagation, 
    \begin{equation}
{\cal T}^{\rm PT}_{c}\asymp e^{- 2 \pi s ( \upsilon - p_0)}
\delta \left[q_0 + 2p_0t+s\ln\left( \frac{v_0^2}{p_0^2}-1\right)-q \right]
  \end{equation}   
The additional term looks like an advance in time, 
violating causality, yet in fact it is an
artifact of the naive semi-classical approximation.

In the opposite limit of a narrow barrier, i.e. small $s$, 
the series converges rapidly. If we set 
$s\rightarrow 0$, $v_0^2s=$const, the $\delta$-function discused in
Section 5 is reproduced. One has no problems with the causality in
that approximation.



\begin{thebibliography}{99}
{\frenchspacing \small
\bibitem{landau-qm}
L.D. Landau and E.M. Lifshitz, "Quantum Mechanics", Pergamon, London,
1958. 
\bibitem{schiff}
L. Schiff, ``Quantum Mechanics", McGraw-Hill, New York, 1968, Ch. 8,9.
\bibitem{buttiker&landauer}
M. Buttiker and R. Landauer, Phys. Rev. Lett. {\bf 49}, 1739 (1982). 
\bibitem{leavens&aers}
C.R. Leavens and G.C. Aers, Phys. Rev. {\bf B39}, 1202 (1989).
\bibitem{hauge&stovneng}
E.H. Hauge and J.A. St{\o}vneng, Rev. Mod. Phys. {\bf 61}, 917 (1989).
\bibitem{olkhovsky&recami}
V.S. Olkhovsky and E. Recami, Phys. Rep. {\bf 214}, 339 (1992).
\bibitem{muga&brouard&sala}
J.G. Muga, S. Brouard and R. Sala, Phys. Lett. {\bf A167},24 (1992).
\bibitem{landauer&martin}
R. Landauer and Th. Martin, Rev. Mod. Phys. {\bf 66}, 217 (1994).
\bibitem{steinberg&chiao}
A. M. Steinberg and R.Y. Chiao, Phys. Rev. {\bf A49}, 3283 (1994).
\bibitem{dsokol}
D. Sokolovski, Phys. Rev. {\bf A52}, R5 (1995).
\bibitem{steinberg}
A.M. Steinberg, Phys. Rev. Lett. {\bf 74}, 2405 (1995).
\bibitem{ranfagni}
A. Ranfagni, D. Mugnai, P. Fabeni, and G. P. Pazzi, 
Appl. Phys. Lett. {\bf 58}, 774 (1991).
\bibitem{ranphys}
A. Ranfagni, D. Mugnai, P. Fabeni, G. P. Pazzi, G. Naletto, and Sozzi,
Physica {\bf B175}, 283 (1991).
\bibitem{ranphyse}
A. Ranfagni, D. Mugnai, P. Fabeni, G. P. Pazzi, G. Naletto, and Sozzi,
Phys. Rev. {\bf E48}, 1453 (1993). 
\bibitem{enders}
A. Enders and G. Nimtz,
Phys. Rev. {\bf B47}, 9605 (1993). 
\bibitem{chiaoetal-exper}
A.M. Steinberg, P.G. Kwiat and R.Y. Chiao, Phys. Rev. Lett. {\bf 71},
708 (1993).
\bibitem{krausz}
Ch. Spielmann, R. Szipocs, A. Sting, and F. Krausz, 
Phys. Rev. Lett. {\bf 73}, 2308 (1994).
\bibitem{winter}
R.G. Winter, Phys. Rev. {\bf 123}, 1503 (1961).
\bibitem{pollak}
N. Abu-Salby, D.J. Kouri, M. Baer and E. Pollak, 
J. Chem. Phys. {\bf 82}, 4500 (1985).
\bibitem{schulmandelta}
B. Gaveau and L.S. Schulman, J. Phys. {\bf A19}, 1833 (1986).
\bibitem{elberfeld}
W. Elberfeld and M. Kleber, Am. J. Phys. {\bf 56}, 154 (1988).
\bibitem{gurvitz}
S.A. Gurvitz, Phys. Rev. {\bf A38}, 1747 (1988).
\bibitem{suen}
W.-M. Suen and K. Young, Phys. Rev. {\bf A43}, 1 (1991).
\bibitem{caldirola-kanai1}
S. Baskoutas and A. Jannusis, J. Phys. {\bf A25}, L1299 (1992).
\bibitem{carvalho}
T.O. de Carvalho, Phys. Rev. {\bf A47}, 2562 (1993).
\bibitem{albeverio}
S. Albeverio, Z. Brzezniak and L. Dabrowski, 
J. Phys. {\bf A27}, 4933 (1994).
\bibitem{kleber}
M. Kleber, Phys. Rep. {\bf 236}, 331 (1994).
\bibitem{moshinsky}
G. Garcia-Calderon, J.L. Mateos, and M. Moshinsky, 
Phys. Rev. Lett. {\bf 74}, 337 (1995).
\bibitem{mclaughlin}
D.W. McLaughlin, J. Math. Phys. {\bf 13}, 1099 (1972).
\bibitem{marpopov}
M. S. Marinov and V. S. Popov, Fortschr. Phys. {\bf 25}, 373 (1977).
\bibitem{sokolovski&baskin}
D. Sokolovski and L.M. Baskin, 
Phys. Rev. A. {\bf 36}, 4604 (1987).
\bibitem{wardlaw}
W. Jaworski and D.M. Wardlaw, 
Phys. Rev. A. {\bf 43}, 5137 (1991).
\bibitem{aoyama}
H. Aoyama and T. Harano, Nucl. Phys. {\bf B446}, 315 (1995).
\bibitem{wigner}
E.P. Wigner, Phys. Rev. {\bf 40}, 749 (1932).
\bibitem{carruthers}
P. Carruthers and F. Zachariasen, Rev. Mod. Phys. {\bf 55}, 245 (1983).
\bibitem{hilleryetal}M. Hillery, R.F. O'Connell, 
M.O. Scully, and E.P. Wigner, Phys. Rep. {\bf 106}, 121 (1984).
\bibitem{balazs&jennings}
N.L. Balazs and B.K. Jennings, Phys. Rep. {\bf 104}, 347 (1984).
\bibitem{peres}
A. Peres, Physica Scripta {\bf 34}, 736 (1986).
\bibitem{botermans}
W. Botermans and R. Malfliet, Phys. Rep. {\bf 198}, 115 (1990).
\bibitem{aichelin}
J. Aichelin, Phys. Rep. {\bf 202}, 233 (1991).
\bibitem{abe&suzuki}
S. Abe and N. Suzuki, Phys. Rev. A. {\bf 45}, 520 (1992).
\bibitem{bonasera}
A. Bonasera, V. N. Kondratyev, A. Smerzi, and E. A. Remler, 
Phys. Rev. Lett. {\bf 71}, 505 (1993).
\bibitem{bizarro}
J. P. Bizarro, Phys. Rev. {\bf A49}, 3255 (1994).
\bibitem{mrowczynski}
S. Mrowczynski, and B. Muller, Phys. Rev. {\bf A50}, 7542 (1994).
\bibitem{royer}A. Royer, Phys. Rev. {\bf A43}, 44 (1991).
\bibitem{ghosh}
S.K. Ghosh and A.K. Dhara, Phys. Rev. {\bf A44}, 65 (1991).
\bibitem{molzahn&osborn}
F.H. Molzahn and T.A. Osborn, Ann. Phys. {\bf 230}, 343 (1994). 
\bibitem{lee}
H.-W. Lee, Phys. Rev. {\bf A50 }, 2746 (1994).
\bibitem{smerzi}
A. Smerzi, Phys. Rev. {\bf A52 }, 4365 (1995).
\bibitem{JACKIW&WOO}R. Jackiw and G. Woo, Phys. Rev.
{\bf D12}, 1643 (1975).
\bibitem{muga&snider}
J.G. Muga and R.F. Snider, Phys. Rev. A. {\bf 45}, 2940 (1992).
\bibitem{poppitz}
T. M. Gould and E. R. Poppitz, Nucl. Phys. {\bf B418}, 131 (1994).
\bibitem{peev}
P. Kasperkovitz and M. Peev, Phys. Rev. Lett. {\bf 75}, 990 (1995).
\bibitem{BERRY}
M. V. Berry, Phil. Trans. Roy. Soc. {\bf 287}, 237 (1977).
\bibitem{LITTLEJOHN}
R. G. Littlejohn, Phys. Rep. {\bf 138}, 193 (1986).
\bibitem{stroud}
M. Mallalieu and C.R. Stroud, Jr., Phys. Rev. {\bf A49}, 2329 (1994).
\bibitem{balazs&voros}
N.L. Balazs and A. Voros, Ann. Phys. {\bf 199}, 123 (1990).
\bibitem{muga-num}
J.G. Muga, J. Phys. A. {\bf 24}, 2003 (1991).
\bibitem{defendi}
A. Defendi and M. Roncadelli, Phys. Lett. {\bf A187}, 289 (1994).
\bibitem{scullyfound}
H.W. Lee and M.O. Scully, Found. of Phys. {\bf 13}, 61 (1983).
\bibitem{jensen}
K.L. Jensen and F. A. Buot, Appl. Phys. Lett. {\bf 55}, 669 (1989).
\bibitem{remler}
E.A. Remler, Ann. Phys. {\bf 95}, 455 (1975).
\bibitem{garraway}
B.M. Garraway and P.L. Knight, Phys. Rev. {\bf A50}, 2548 (1994).
\bibitem{marin91}
M. S. Marinov, Phys. Lett. {\bf A153}, 5 (1991).
\bibitem{analy}  
M. S. Marinov and B. Segev, 
``Analytical properties of scattering amplitudes
in one-dimensional quantum theory'', J. Phys. A (to be published).
\bibitem{abram}  
``Handbook of Mathematical Functions", eds.
M. Abramowitz and I. A. Stegun, NBS, 1964.
\bibitem{time}  
M. S. Marinov and B. Segev, ``On the concept of the tunneling time", 
Technion, Haifa 1996. 
\bibitem{slater}
L. J. Slater, ``Generalized Hypergeometric Functions", Cambridge 
University Press, 1966.
}   

\end{thebibliography}
\end{document}